\documentclass[aps,amsmath,amssymb,prb,preprint]{revtex4-2}
\usepackage{graphicx,subfigure,color,tikz}

\newcommand{\bfg}{\mathbf{g}}
\newcommand{\bft}{\mathbf{t}}
\newcommand{\bfT}{\mathbf{T}}
\newcommand{\tr}{\text{tr}}
\newcommand{\bfs}{\mathbf{\sigma}}

\begin{document}

\title{\textbf{Bias controlled Interlayer Exchange Coupling} 
}

\author{Nathan A. Walker, Alex D. Durie and Andrey Umerski}
\affiliation{
 Department of Mathematics and Statistics, Open University, Milton Keynes, MK7 6AA, U.K.
}

\date{\today}

\begin{abstract}
    We demonstrate, using computer simulations and a non-equilibrium Greens function approach, that the sign of the out-of-equilibrium interlayer exchange coupling (ooeIEC) can change in the presence of an externally applied electrical bias. Our system consists of an insulating section connected to an exchange coupled ferromagnetic (FM) tri-layer, sandwiched between semi-infinite leads. When the exchange coupled trilayer contains a quantum-well state confined in the hybridisation gap (HG) of the FM, we find that a relatively small applied electrical bias can switch the lowest energy state of the tri-layer between parallel (P) and anti-parallel (AP) configurations. We consider three cases for the insulating section; a single tunnelling barrier, a resonant tunnelling barrier and an amorphous insulating barrier and, in each case, show that the bias dependence of the ooeIEC is strongly dependent on the system conductance. We find that the lowest switching current densities are achieved with strongly confined quantum well states.
\end{abstract}

\maketitle

%%%%%%%%%%%%%%%%%%%%%%%%%%%%%%%%%%%%%%%%%%%%%%%%%%%%%%%%%%%%%%%%%%%%%%%%%%%%

\section{Introduction}
Due to its non-volatility and high read/write speeds, magnetoresistive random-access memory (MRAM) \cite{MRAM1,MRAM2} is one of the principal contenders for a universal memory, someday replacing the separate SSD/HDD and RAM memory devices used in modern computers with one, single, memory architecture \cite{MRAMgood}. Not only does MRAM have the potential to increase the speed and reduce the energy consumption of ‘everyday’ computers, it is also desirable for use in next generation technology such as neuromorphic systems \cite{neuro_mentions_MRAM,neuro_review}.

The key component of MRAM is the magnetic tunnel junction (MTJ) \cite{MTJreview}, composed of two ferromagnetic (FM) leads separated by an insulating barrier, typically several layers of crystalline MgO. The magnetic moments of the FM leads can be arranged in a parallel (P) or anti-parallel (AP) configuration. It has been found \cite{ikeda2008} that the electrical resistance of the junction can be up to 6 times higher at room temperature in the AP configuration, making it ideal for storing and reading information. 
However, a major obstacle to the development of commercially viable MRAM has been the difficulty in achieving an energy efficient writing mechanism, that is the ability to switch between P and AP configurations. Some of the most researched writing mechanisms, spin-transfer torque (STT) \cite{STT1,STT_recent_review}, spin-orbit torque (SOT) \cite{SOT1}, and more recently orbit-hall torque \cite{orbitcurrentdownside1,orbitcurrentdownside2}, to-date require relatively high current densities \cite{RecentreviewhighSTTcurrent,SOTdownside}.  It is therefore pertinent to explore alternative methods of magnetic switching which minimise the charge current required.

In this communication we propose a different approach based on interlayer exchange coupling (IEC).
It is well known \cite{Parkin} that, for transition metal spacers, the IEC of a FM/non-magnetic spacer(NM)/FM tri-layer oscillates as a function of NM thickness and a handful of attempts by various groups (experimental and theoretical) have been made to take advantage of this property of IEC with the goal of achieving magnetic switching.

An early theoretical work by You and Suzuki \cite{YouSuzuki} considers a FM/NM/FM IEC junction coupled to a semiconductor.  Applying a bias alters the height and thickness of the Schottky barrier, thus modifying the reflection coefficients and leading to a change of sign of the IEC. However, these calculations were for equilibrium IEC in the absence of bias, and required rather specific model parameters to achieve an effect.

Bias dependence of IEC was discussed for several systems in a work by Haney \textit{et. al.} \cite{Stiles} using a one-dimensional, single orbital, tight-binding, non-equilibrium Greens function approach.
They find that for fully metallic systems, with large spacer thickness, the IEC is a weak oscillatory function of the applied bias but would require unfeasibly large currents to induce switching.
For IEC junctions with an insulating spacer they find that the IEC depends quadratically on the applied bias and so cannot give rise to switching.

A method in which a ferroelectric layer is coupled to a FM/NM/FM junction was considered by Fechner \textit{et}. \textit{al}., \cite{Fechneretal}. Fully realistic calculations show that the IEC changes sign with a change of direction of electric polarisation, which could be achieved by an external electric field.  The polarisation is characterised by the displacement of oxygen atoms within the ferroelectric, and the authors propose a change of reflection coefficients at the FM/ferroelectric interface as the likely mechanism for the IEC sign change.

% In another work by by Sayed \textit{et. al} \cite{litrevrecent2}, a model single-orbital system with a resonant tunnel barrier replacing the NM spacer of an IEC tri-layer was discussed. 
% A bias applied to the junction allows for resonant tunnelling across the spacer, apparently resulting in an IEC that oscillates with respect to bias. However, in this system the equilibrium exchange coupling is negligible, suppressed by large barriers, so it is unclear whether this lack of stability makes the system suitable for application. 

On the experimental side a few groups have either attempted or reported to observe switching of IEC in a variety of settings. 
%The group of Lavrijsen \cite{Lavrijsen} investigated the influence of a strong applied electric field on the IEC. They considered an IEC junction coupled to a dielectric, MgO insulating layer, but found that they were unable to observe any effect using very strong electric field strengths up to 0.5Vnm$^{-1}$. 

Perhaps the first experimental demonstration of voltage control of IEC was reported by Newhouse-Illige \textit{et. al.} \cite{extraexperiment1}. Reversible and deterministic switching of the sign of the IEC of a perpendicular magnetic tunnel junction consisting of CoFeB FM's and an amorphous Gd$\text{O}_x$ barrier was demonstrated. The effect was attributed to the ability to move oxygen vacancies within the Gd$\text{O}_x$ tunnel barrier and a large induced net magnetic moment of the Gd ions.  While switching was only observed at high temperatures and over tens of seconds, this work represents a crucial proof-of-concept for voltage-controlled IEC.

More recently, Zhang \textit{et}. \textit{al}. \cite{litrevrecent1} demonstrated electric-field switching of a composite FM/NM/FM IEC junction and an MgO MTJ.  They report a reversible magnetisation switching effect dominated by IEC with negligible STT. The current density per write operation (magnetic switch) is measured to be one order of magnitude lower than the best reported STT devices.
They attribute the effect to biased induced changes in spin-dependent reflectivities at the MgO/FM interface and consider a single-orbital parabolic band model and Bruno reflection theory to justify their claim.  However, their model is for weakly confined states without quantum well confinement, and hence the calculated IECs are some 100 times smaller than the measured ones.  
When realistic potentials are used using this theory, our calculations show that the change in IEC due to bias, using the proposed mechanism, is at least two orders of magnitude too weak to cause switching and so the theory cannot explain the observed phenomena.

\begin{figure}[h]
    \centering
    \includegraphics[width=0.7\linewidth]{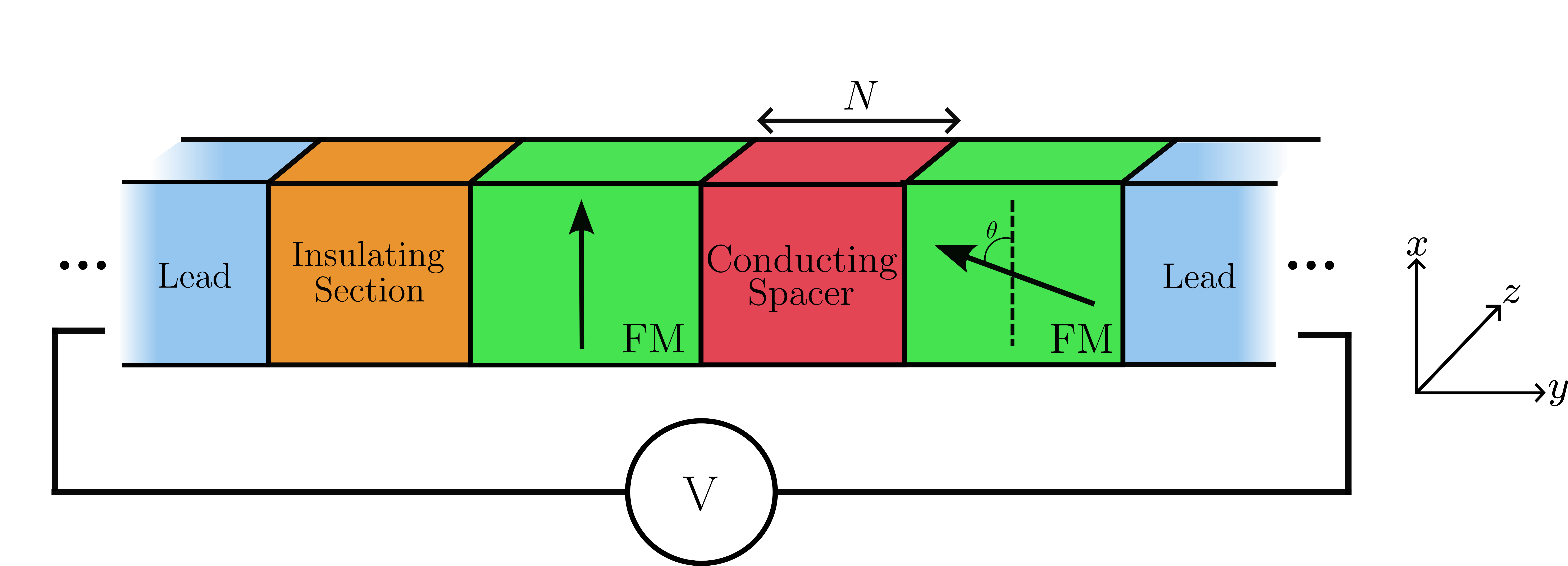}
    \caption{The multilayer system considered in this communication, consisting of an exchange coupled tri-layer (thickness $N$) and insulating section, sandwiched between two non-magnetic, semi-infinite leads. }
    \label{fig:fig 14}
\end{figure}

With the exception of Ref.\cite{Stiles} (which concludes that switching of IEC requires unfeasibly large currents), all the works discussed above employ spin-dependent changes in the reflectivities as the main mechanism for effecting the IEC. 
In this communication we explore an alternative non-equilibrium theory, which can induce switching even in strongly coupled, quantum well, IEC systems, and may well be the true mechanism underlying the results of Zhang \textit{et}. \textit{al}.
We consider multilayer systems of the form shown in FIG. \ref{fig:fig 14},  consisting of an exchange coupled FM/NM/FM trilayer connected to an insulating section and NM leads, under the influence of a finite applied electrical bias, $V$.  The purpose of the barrier is simply to maintain a finite bias, and we consider 3 different types: a single barrier, double (resonant tunnel) barrier and an amorphous insulating barrier. In each case we calculate the effect of the applied bias on the out-of-equilibrium IEC (ooeIEC).

We are interested in IEC junctions where there is quantum well confinement in the spacer for one of the electron spin states.  This is the case for most real systems exhibiting strong oscillatory IEC, such as CoRu, CoCu, FeAu and FeAg.  For these systems quantum well confinement occurs because of the presence of a hybridisation gap (HG) in the band structure of the corresponding spin state of the FM at the Fermi level.
In previous works on altering the IEC, the presence of the HG has been ignored -- except for being crudely modelled by a single-band, half-metal.
However, in addition to strong oscillatory IEC, the presence of the HG leads to an interesting effect, crucial to our calculations --- it makes the IEC highly sensitive to the position of the Fermi-level within the gap. 
As discussed in Section~\ref{sec:motiv}, we take advantage of this effect by using a relatively small applied bias across the IEC junction to engage states throughout the HG in the ooeIEC.
As an exemplar of such quantum well IEC systems, in this communication we choose Co for the FM layer and Cu for the NM layers.
Co/Cu/Co (001) IEC junctions have been extensively studied in the literature and are known to exhibit strong oscillatory IEC~\cite{devries95,weber1995,qiu2002,mathon1995} due to the presence of a HG in the band structure of minority spin Co.
However we emphasise that there is nothing particularly special about the Co/Cu system --- it is a typical IEC system, and our findings in this communication are applicable to most quantum well IEC systems.

The outline of this communication is as follows.
In Section~\ref{sec:II}, we set up the necessary mathematical background to define our multilayer junction, show how we determine the ooeIEC in terms of the parameters of the system and briefly discuss how the calculations are performed.  Section~\ref{sec:motiv} describes our motivation for studying this system and introduces further details of our tight-binding parameters.  Section~\ref{sec:results} is the main part of this communication and is split into three subsections, each of which presents our results and conclusions for the different insulating sections.  Finally we summarise our results in the conclusion.

To simplify our calculations, in this communication we consider a 2-band model which mimics the band structure of FCC Co for the FM and FCC Cu for the NM spacer and leads.  
Our results are calculated in computer simulations using a tight-binding model and non-equilibrium Landauer approach.

%%%%%%%%%%%%%%%%%%%%%%%%%%%%%%%%%%%%%%%%%%%%%%%%%%%%%%%%%%%%%%%%%%%%%%%%%%%%

\section{Out-of-Equilibrium Interlayer Exchange Coupling}\label{sec:II}

We consider the INS/FM/NM/FM multi-layer system discussed above and initially assume the system grows perfectly epitaxially so there is in-plane translational symmetry. As such, we express the Hamiltonian in the mixed-space basis $ \{ \boldsymbol{k_{\parallel}},n \} $ where $\boldsymbol{k_{\parallel}}=(k_x,k_z)$ is the in-plane wave vector and $n$ is the real-space plane index. We use a tight-binding formalism with on-site potentials $\textbf{u}_{n\nu\sigma,n\nu'\sigma'}(\boldsymbol{k_{\parallel}})$ and inter-planar hoppings $\textbf{t}_{n\nu\sigma,n'\nu'\sigma'}(\boldsymbol{k_{\parallel}})$ where $\nu$ is the orbital index and $\sigma$ is the spin index.

We define $\theta$ as the angle between the magnetic moments of left and right-hand FM layers and assume that the magnetic moments lie in-plane (see FIG. \ref{fig:fig 14}). Furthermore, we assume that the local potentials do not change in going from the P to AP configurations. Then, in the presence of an applied electrical bias, the ooeIEC per atom ($J$) is defined as the work required to rotate the right FM's magnetic moment through an angle $\theta =0$ to $\pi$. That is, 
\begin{equation}
    \label{eqn:eqn 1}
    J=\int_{0}^{\pi} d\theta \ \tau_{y}
\end{equation}
\noindent where $\tau_{y}(\theta)$ is the $y$-component of the torque exerted on the right-hand FM's magnetic moment. As described in Ref.~\cite{MathonUmerskiReview}, the torque is equal to the total rate of change of angular momentum of precessing electrons which, by continuity, is equal to the rate of flow of the $y$-component of the spin angular momentum (spin-current) across the spacer, $j(\epsilon,\textbf{k}_{\parallel},\theta)$. 

To calculate the non-equilibrium spin-current in the presence of an applied bias we use the Landauer formalism \cite{Landauer} and assume that electrons of all possible momenta, energy and spin orientation are injected into the system to write equation (\ref{eqn:eqn 1}) as
\begin{equation}
    \label{eqn:eqn 2a}
     J=\frac{1}{A_{BZ}} \int_{0}^{\pi} d\theta \int_{BZ} d^2 \boldsymbol{k}_{\parallel} \int_{-\infty}^{\infty} d\epsilon \ \left(f_Lj_j+f_Rj_R\right)
\end{equation}
\noindent where $A_{BZ}$ is the area of the 2D Brillouin zone parallel to the layers, $j_{L/R}$ are the out-of-plane spin-current densities originating from the left/right leads and $f_{L/R}=f(\epsilon-\mu_{L/R})$ are the Fermi functions at temperature $T$ with $\mu_{L/R}$ the chemical potentials of the left/right leads respectively.

It can be shown that, between any two NM spacer layers $n$ and $n+1$, $j_{L}$ has the following simple expression
\[j_L=4\,\tr \left(\bfT.\text{Im} (\bfg_{L}) .\bfT^{\dagger}.\text{Im} (\bfs_y \bfg_{R}) \right) \quad \text{where}\quad \bfT=(1-\bft^{\dagger}\bfg_L\bft\bfg_R)^{-1}\bft^{\dagger}.\]
Here $\bfs_y$ is the $y$-component Pauli matrix, $\bfg_{L/R}$ are the left and right {\emph surface} Greens functions if the system were cleaved between layers $n$ and $n+1$, $\bft$ is the hopping between these layers and $\text{Im} (\bfg)=\frac{1}{2i}(\bfg-\bfg^{\dagger})$.
A similar expression for $j_R$ is obtained from $j_L$ by exchanging $\bfg_L \leftrightarrow \bfg_R$ and $\bft \leftrightarrow \bft^{\dagger}$ throughout.  These expressions are fully equivalent to those derived using, the Keldysh non-equilibrium Greens function formalism \cite{edwards2006}.

For calculation purposes it is useful to express Equation~(\ref{eqn:eqn 2a}) in the following symmetrised form
\begin{equation}
    \label{eqn:eqn 2}
     J=\frac{1}{2A_{BZ}} \int_{0}^{\pi} d\theta \int_{BZ} d^2 \boldsymbol{k}_{\parallel} \int_{-\infty}^{\infty} d\epsilon \ \left[(f_L+f_R)(j_L+j_R)+(f_L-f_R)(j_L-j_R)\right].
\end{equation}
We refer to the first term here as the IEC term and the second the spin-STT term.  While both terms depend on the applied bias, it is only the IEC term that contributes at zero bias, at which point it becomes identical to the IEC discussed in Section I. 
It should be noted however that, while this form is mathematically useful, for out-of equilibrium systems like those considered here it makes little physical sense to consider either term in isolation.

The integrand of Equation (\ref{eqn:eqn 2}) is highly singular in both energy and momentum and so the integral is numerically intensive. However, for the case where the right-hand FM of FIG.~\ref{fig:fig 14} is rotated, it can be shown that the integrand of the IEC term can be expressed as 
\[j_L+j_R = -4\frac{d\ }{d\theta}\,\text{Im}\left(\ln \det \bfT\right).\]
If the left-hand FM is rotated, the sign changes.  
So the $\theta$ integral can be performed exactly and, since $\bfT$ is analytic in energy, the energy integral can be evaluated in the complex plane using the method of Matsubara \cite{MatsubBook}. On the other hand, the STT term is not an exact derivative of $\theta$, and is non-analytic in energy, so fully numerical methods must be used to evaluate it. However, in this work we consider only relatively small bias (of the order of the HG width) and so the domain of energy integration, dictated by the Fermi function factor $f_L-f_R$ (the difference between the left and right chemical potentials), is also small, reducing computational overhead significantly.  Nevertheless, even for such small bias, typically hundreds of integrand evaluations are required for convergence in energy alone.

All results were derived using an adaptive integration algorithm for the energy integral, a weighted sum over a grid of 1540 points in the irreducible (simple cubic) Brillouin zone for the $\textbf{k}_{\parallel}$ integral and Simpsons rule for the $\theta$ integral. Layer thicknesses refer to the number of atomic planes throughout, and, unless specified otherwise, FM layer thicknesses are fixed at 5, $T=316K$, the equilibrium (right lead) chemical potential is $\mu_R=0$, the left lead chemical potential is $\mu_L=eV$.

%%%%%%%%%%%%%%%%%%%%%%%%%%%%%%%%%%%%%%%%%%%%%%%%%%%%%%%%%%%%%%%%%%%%%%%%%%%%%%%%%%%%%%%

\section{Motivation and Model}\label{sec:motiv}

\begin{figure}[h]
    \centering
    \includegraphics[width=0.7\linewidth]{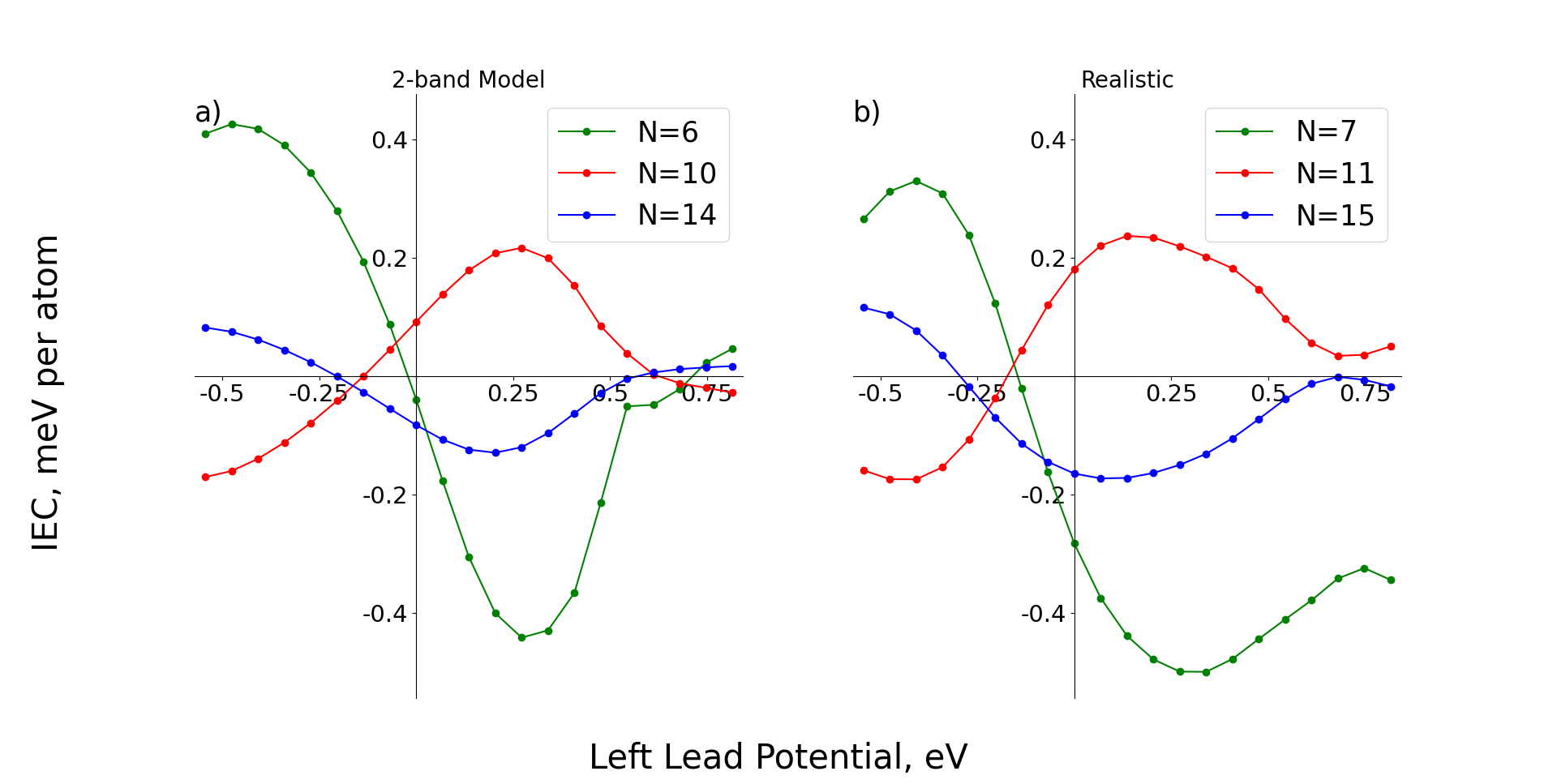}
    \caption{Comparison of IEC versus left lead potential for (a) 2-band model and (b) fully realistic, calculation of a Co/Cu(N)/Co tri-layer, with potential ranging across the Co HG. }
    \label{fig:fig 3}
\end{figure}

We are motivated to study the effect of bias on the ooeIEC following a theoretical result, by one of us, \cite{UmerskiPhase} which demonstrates that very small changes in the chemical potential of the FM can cause the IEC to switch between P and AP configurations. This phenomenon has been observed experimentally by doping the FM layers in a Co/Ru/Co IEC junction \cite{ebels1998}.  The effect occurs when the oscillatory IEC is dominated by quantum well states, bound in the spacer because the chemical potential of the FM lies in a HG.
 This is the case for many realistic systems such as CoRu, CoCu, FeAu and FeAg exhibiting strong oscillatory IEC.  The effect is demonstrated in FIG.~\ref{fig:fig 3}, which shows the IEC as a function of the chemical potential in the left-hand lead, for a 2-band model (FIG.~\ref{fig:fig 3}(a)) and fully realistic (FIG.~\ref{fig:fig 3}(b)) calculation of a Co/Cu(N)/Co trilayer.   The IEC of different spacer thicknesses $N$ are shown, and we observe that the IEC consistently flips sign, corresponding to a change in the system configuration (P or AP), as we move the (left) FM potential across the Co HG located at $-0.55 \lesssim \epsilon \lesssim 0.7$ eV.
 This effect is largely independent of the detailed parameters of the system --- occurring in any system where the spacer has quantum well states and the chemical potential moves across the HG of the FM.

\begin{figure}[h]
    \centering
    \includegraphics[width=0.7\linewidth]{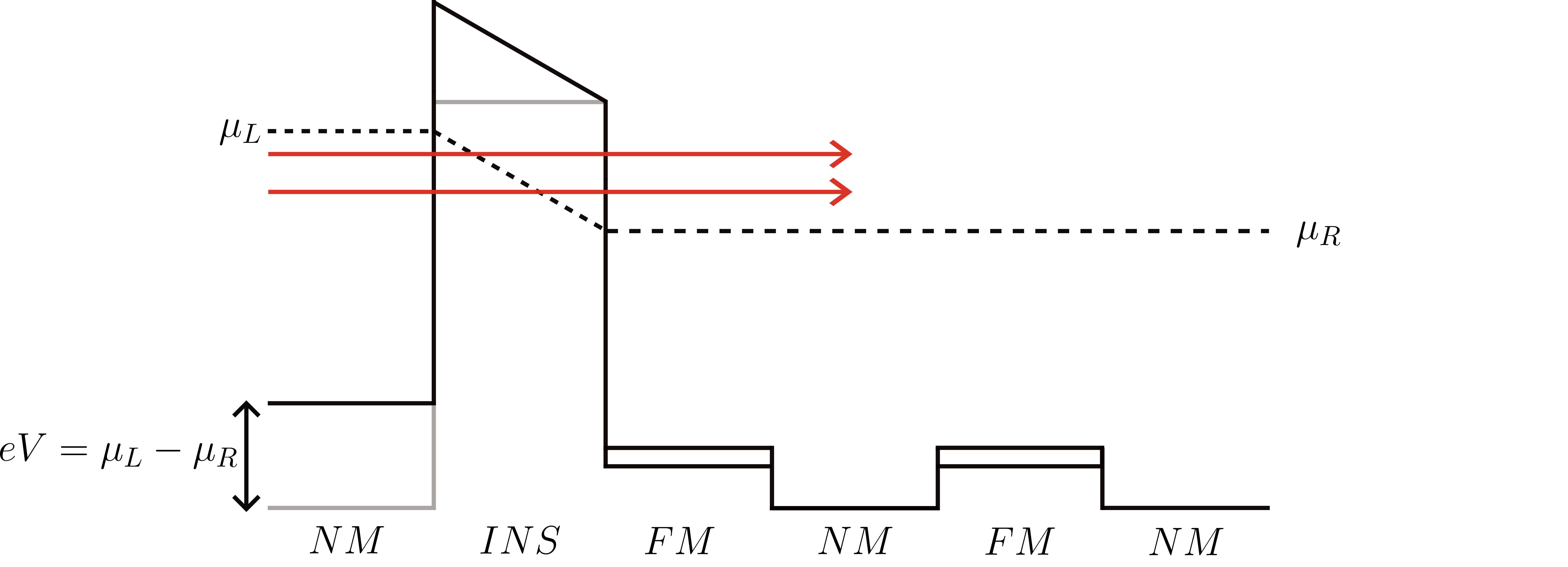}
    \caption{1D potential schematic representing our multilayer system under an applied bias, illustrating how additional states below the left lead chemical potential contribute to the ooeIEC.  The
grey lines represent the system in the absence of a bias. }
    \label{fig:fig 15}
\end{figure}

While the chemical potential of a real FM cannot be altered directly, states above and below the chemical potential can be accessed through the application of a finite applied bias as shown in FIG.~\ref{fig:fig 15}. 
The figure shows a quantum well IEC trilayer, of the type discussed above, connected to an insulating barrier and NM leads.  The potential drop across the barrier illustrates that, by applying positive and negative biases, states above and below the chemical potential in the FM contribute to the ooeIEC.  Furthermore, because HG widths can be small ($\sim 1$eV), relatively small biases can access all states in a gap of the FM.
The aim of this paper is to explore the effect of such an applied bias.

\section{Results for 2-band model}
\subsection{2-band Potentials}\label{sec:potent}
\begin{figure}[h]
    \centering
    \includegraphics[width=0.7\linewidth]{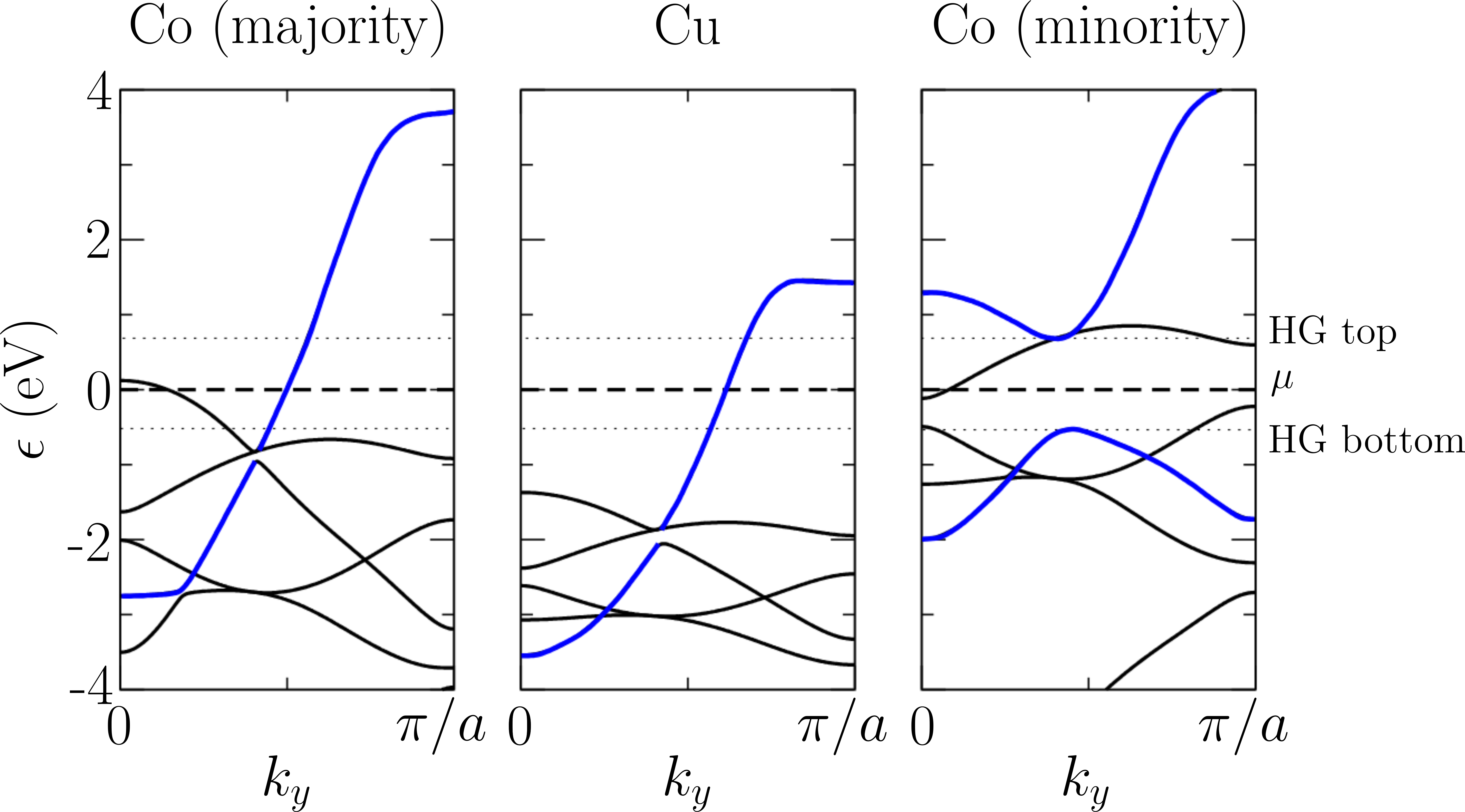}
    \caption{Realistic Co, Cu band structures at the neck of Cu.  With the bands fitted by the 2-band model at the $\Gamma$-point, indicated in blue.  The HG top and bottom are indicated by the dotted lines.}
    \label{fig:fig 13}
\end{figure}

Our tight-binding parameters, shown in Table~\ref{tab:table1}, are obtained by fitting the 2-band parameters at the $\Gamma$-point to relevant realistic FCC Co and Cu bands in the region of the neck of Cu, located at $\boldsymbol{k}_{\parallel}a \approx (\pm0.8\pi,\pm0.8\pi)$ where $a=3.6$\AA~is the lattice constant of Cu.  This region of the Brillouin zone is the dominant source of oscillatory IEC in Co/Cu/Co trilayers, and the 2-band parameters replicate the minority-Co $sd$-like HG, and the Cu and majority-Co $s$-like bands crossing the Fermi level. 
The realistic band structure of Co and Cu at the neck is shown in FIG.~\ref{fig:fig 13}.  The blue bands are those that the 2-band model are fitted to, and the dotted lines indicate the HG top and bottom (see Ref.~\cite{durie22} for details of the fitting procedure).

% \begin{table}[h]
% \caption{Tight binding parameters for Co and Cu}\label{tab:table1}%
% \begin{tabular}{|c|c|c|c|c|c|}
% \hline
%   & $u_{ss}$ & $u_{sd}$ & $t_{ss}$ & $t_{sd}$ & $t_{dd}$\\
% \hline
% \ Cu\ & $-0.6006$ &\ 0.5140\ &\  0.0621\ &\ 0.0244\ & $-0.1609\ $\\
% \hline
% \ Co$^\uparrow$\ & $-0.3392$ & 0.6533 & $-0.00752$ & $-0.0410$ & $-0.1811$\\
% \hline
% \  Co$^{\downarrow}$\ & $-0.2237$ & 0.4018 & 0.0497 & 0.0101 & $-0.0867$\\
% \hline
% \end{tabular}
% \end{table}

\begin{table}[h]
\caption{Tight binding parameters for Co and Cu}\label{tab:table1}%
\begin{tabular}{|c|c|c|c|c|c|}
\hline
  & $u_{ss}$ & $u_{dd}$ & $t_{ss}$ & $t_{sd}$ & $t_{dd}$\\
\hline
\ Cu\ & $-8.1719$ & $6.9929$ & $0.8446$ & $0.3320$ & $-2.1896$\\
\hline
\ Co$^\uparrow$\ & $-4.6151$ & $8.8886$ & $-0.1023$ & $-0.5578$ & $-2.4640$\\
\hline
\  Co$^{\downarrow}$\ & $-3.0437$ & $5.4675$ & $0.6763$ & $0.1376$ & $-1.1790$\\
\hline
\end{tabular}
\end{table}

\begin{figure}[h]
    \centering
    \includegraphics[width=0.7\linewidth]{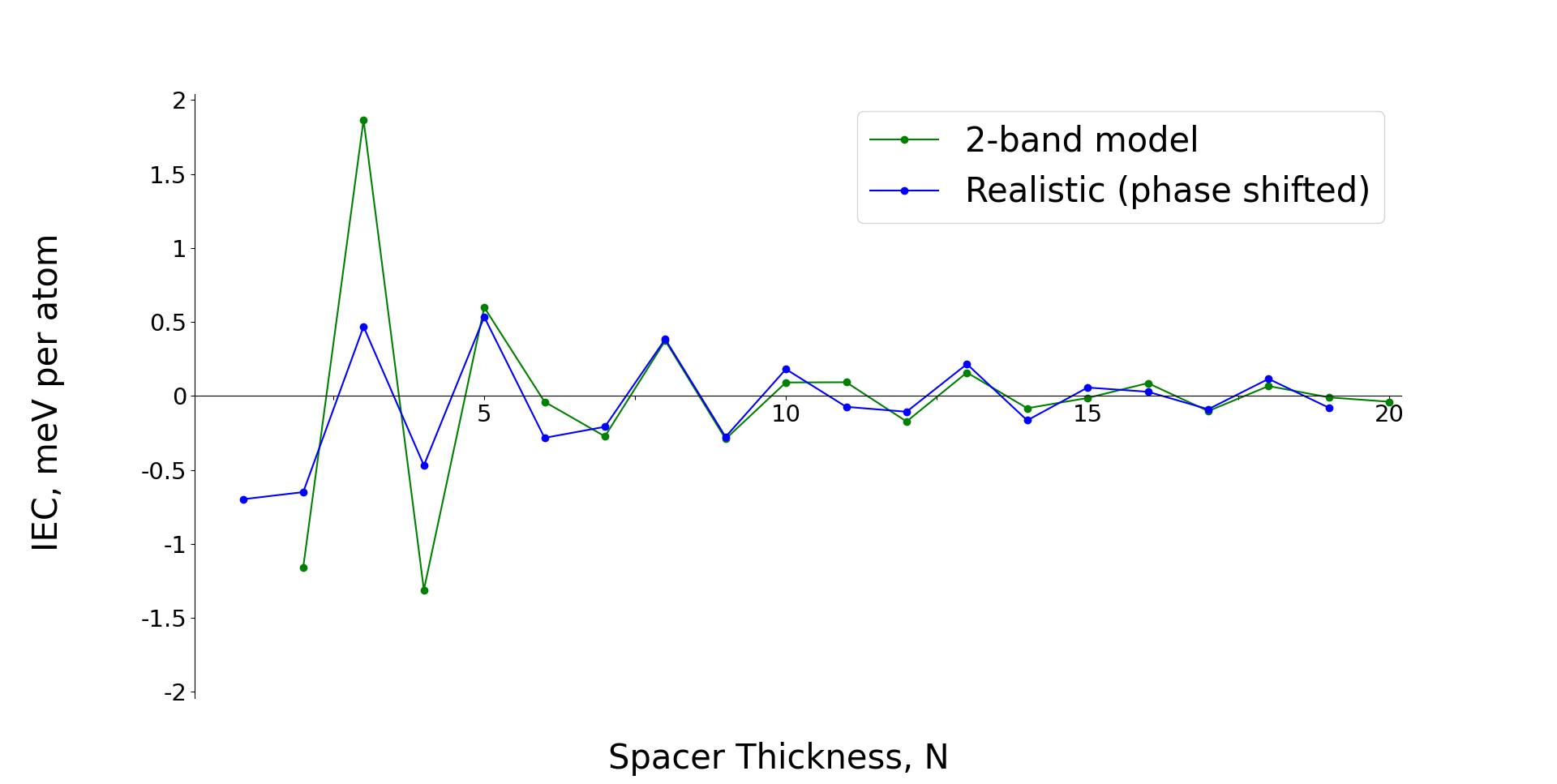}
    \caption{Comparison of IEC vs spacer thickness for 2-band and realistic models for a FM/NM(N)/FM tri-layer. The IEC for the realistic calculation is shifted by 1 layer to the left for ease of comparison.}
    \label{fig:fig 2}
\end{figure}

These 2-band parameters lead to the strong quantum well confinement of minority carriers, and weak confinement of majority carriers, responsible for strong oscillatory IEC observed for realistic Co/Cu/Co.  FIG.~\ref{fig:fig 2} compares the IEC calculated from the 2-band model parameters and those of a fully realistic 9-band calculation.  Note that for ease of comparison, the realistic calculation has been shifted to the left by 1~layer.
 We observe that, apart from the overall phase difference of about 1 layer, the period of oscillation and amplitude of both calculations match well, especially for thicker spacers. 

FIG.~\ref{fig:fig 3} compares the model and realistic calculated IEC of Co/Cu($N$)/Co as a function of the potential in the left lead, for $N=6$, 10 and 14 in the 2-band model, and for $N=7$, 11 and 15 in the realistic calculation.  Again, except for the difference in spacer thickness $N$,  we observe very good overall agreement with very similar qualitative and quantitative behaviour.  We conclude that our model parameters provide a good approximation for calculating the IEC, in the region of the Fermi level, for FCC Co/Cu/Cu and similar quantum well systems.

The remainder of this communication is devoted to exploring the calculation of ooeIEC in systems of the the form: Cu/INS/Co/Cu/Co/Cu where INS refers to one of three insulating sections: single barrier, double barrier or amorphous barrier.

%%%%%%%%%%%%%%%%%%%%%%%%%%%%%%%%%%%%%%%%%%%%%%%%%%%%%%%%%%%%%%%%%%%%%%%%%%%%%%%%%%%%%%%%%%%

\section{Results}\label{sec:results}

\subsection{Single Barrier}

\begin{figure}[h]
    \centering
    \includegraphics[width=0.7\linewidth]{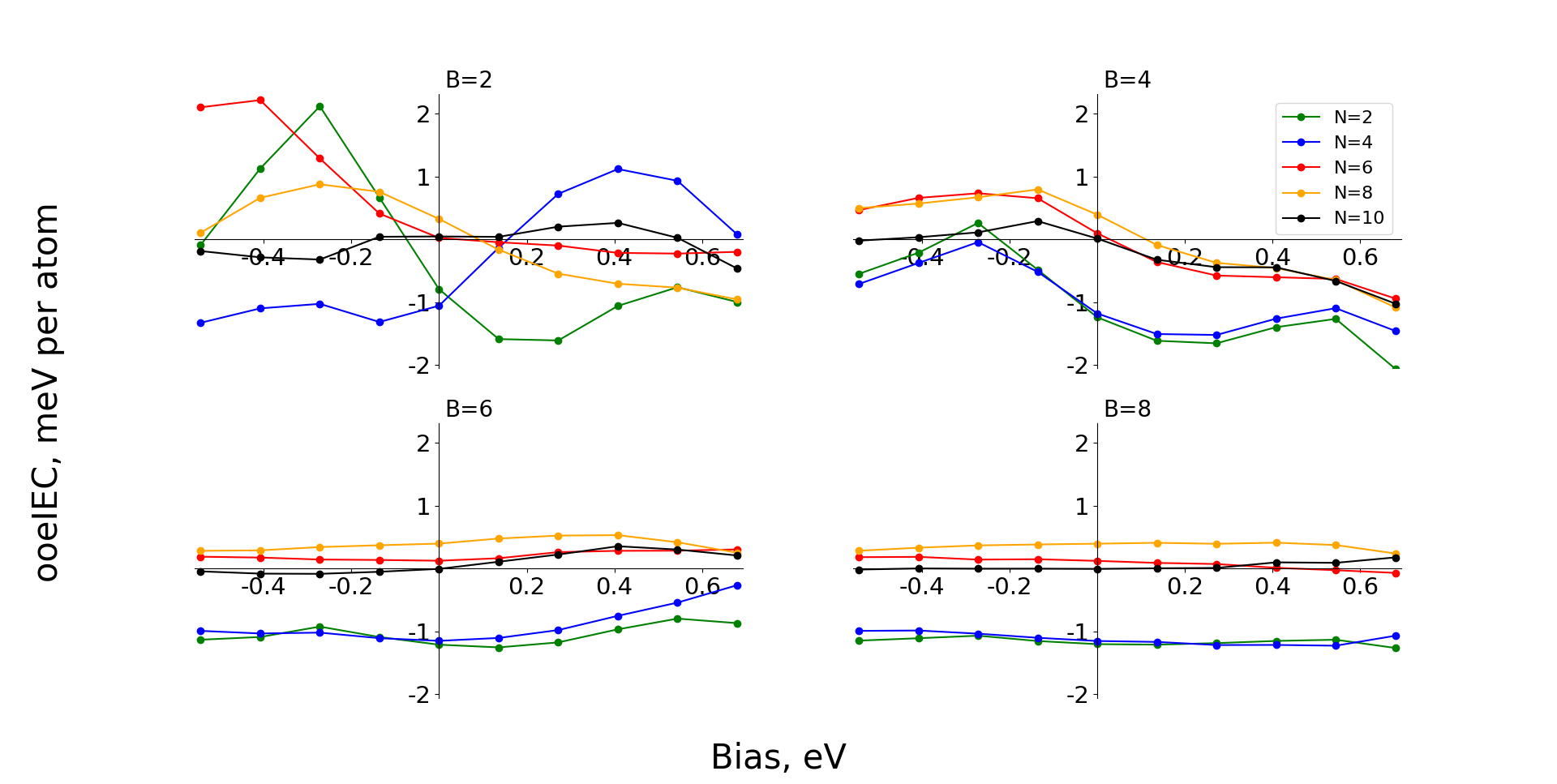}
    \caption{ooeIEC versus bias for a single barrier system of thickness $B=2$ (top left), 4 (top right), 6 (bottom left), 8 (bottom right). }
    \label{fig:fig 6}
\end{figure}

In this section we consider systems of the form Cu/INS($B$)/Co(5)/Cu($N$)/Co(5)/Cu, where $B$ and $N$ refer to the thicknesses of the insulating barrier and Cu spacer respectively.  The insulator is modelled, in the first instance, by the minority Co band structure depicted in FIG.~\ref{fig:fig 13}, which has a gap for energies in the region $-0.55 \lesssim \epsilon \lesssim 0.7$ eV.  
The bias is assumed to drop linearly over the insulating section, and results for the ooeIEC as a function of applied bias, $-0.55 \le eV \le 0.7$ eV, with $B=2,4,6,8$ are shown in FIG.~\ref{fig:fig 6}.

For barrier thickness $B=2$ we observe a very strong non-linear bias dependence as we move across the HG, analogous to that seen in FIG~\ref{fig:fig 3}.  The IEC consistently switches sign for all spacer thicknesses -- corresponding to a flip from parallel to antiparallel magnetic configuration.  
For $B=4$ the effect is lessened slightly, but the bias dependence is still sufficient to change the sign for spacer thicknesses $N=6$, 8 and 10.  For thinner spacers, $N=2$ and $N=4$, the bias dependence is not quite strong enough to demonstrate a change of sign without a constant applied external magnetic field.
 We estimate that barrier thickness $B=4$ corresponds to a charge current density of about $10^8$A/cm$^2$ which, although almost a factor of 100 larger than the best reported STT devices, overcomes a switching magnet which is strongly pinned due to IEC. 

\begin{figure}[h]
     \centering
     \includegraphics[width=0.7\linewidth]{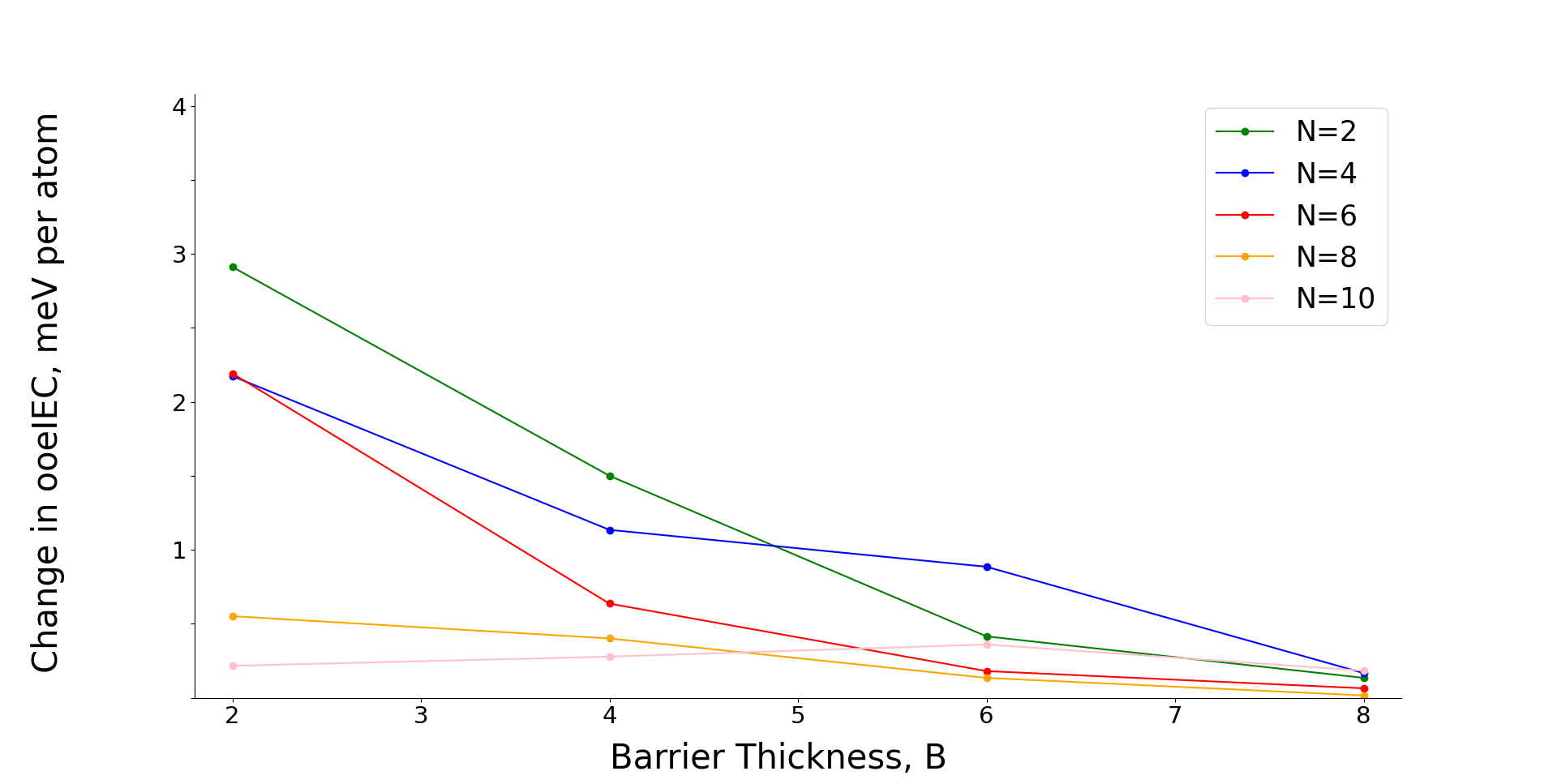}
     \caption{Maximum change in ooeIEC (as a function of bias) for the single barrier system against barrier thickness $B$.}
     \label{fig:fig 23}
\end{figure}

For thicker insulating sections $B=6$ and $B=8$, the ooeIEC becomes largely independent of the bias,  which is to be expected on physical grounds, since in the limit of very thick barriers, the Co/Cu/Co trilayer becomes effectively cut-off from the left-hand lead. 
The barrier thickness dependence is displayed in Fig. \ref{fig:fig 23}, showing that the
amplitude of the effect (the difference between the maximum ooeIEC and the equilibrium
IEC) declines, fairly uniformly, as barrier thickness increases.  

There are two important points to note about these results, which are also applicable to all systems considered in this communication.  Firstly, the contribution from just the change in barrier potential on the equilibrium IEC, is at least one to two orders of magnitude too small to cause switching and cannot account for the effect seen here.  
Secondly, the contributions from the STT and IEC terms in Equation~(\ref{eqn:eqn 2}) are large with significant variation in bias, for all barrier thicknesses -- emphasising the fact that it makes little physical sense to consider either term in isolation.

\subsection{Effect of FM HG width}

Although all quantum well oscillatory exchange coupling effects originate from states confined within a HG, different ferromagnetic materials have different HG widths.  Additionally since the oscillatory IEC originates from those parts of the Brillouin zone where the spacer has a Fermi surface extrema, then different spacers pick out different points in the Brillouin zone and hence experience different HG widths. For example, oscillatory IEC in Co/Cu/Co stems primarily from a HG, in minority Co, of width $\sim 1.25eV$ located at the neck of Cu, whereas IEC in Fe/Au/Fe stems from HGs, in minority Fe, of widths $1.5$eV and $4.5$eV located at the belly and neck of Au respectively. 
So it is pertinent to consider the effect of the FM HG width on the the ooeIEC bias dependence.

\begin{figure}[h]
    \centering
    \includegraphics[width=0.7\linewidth]{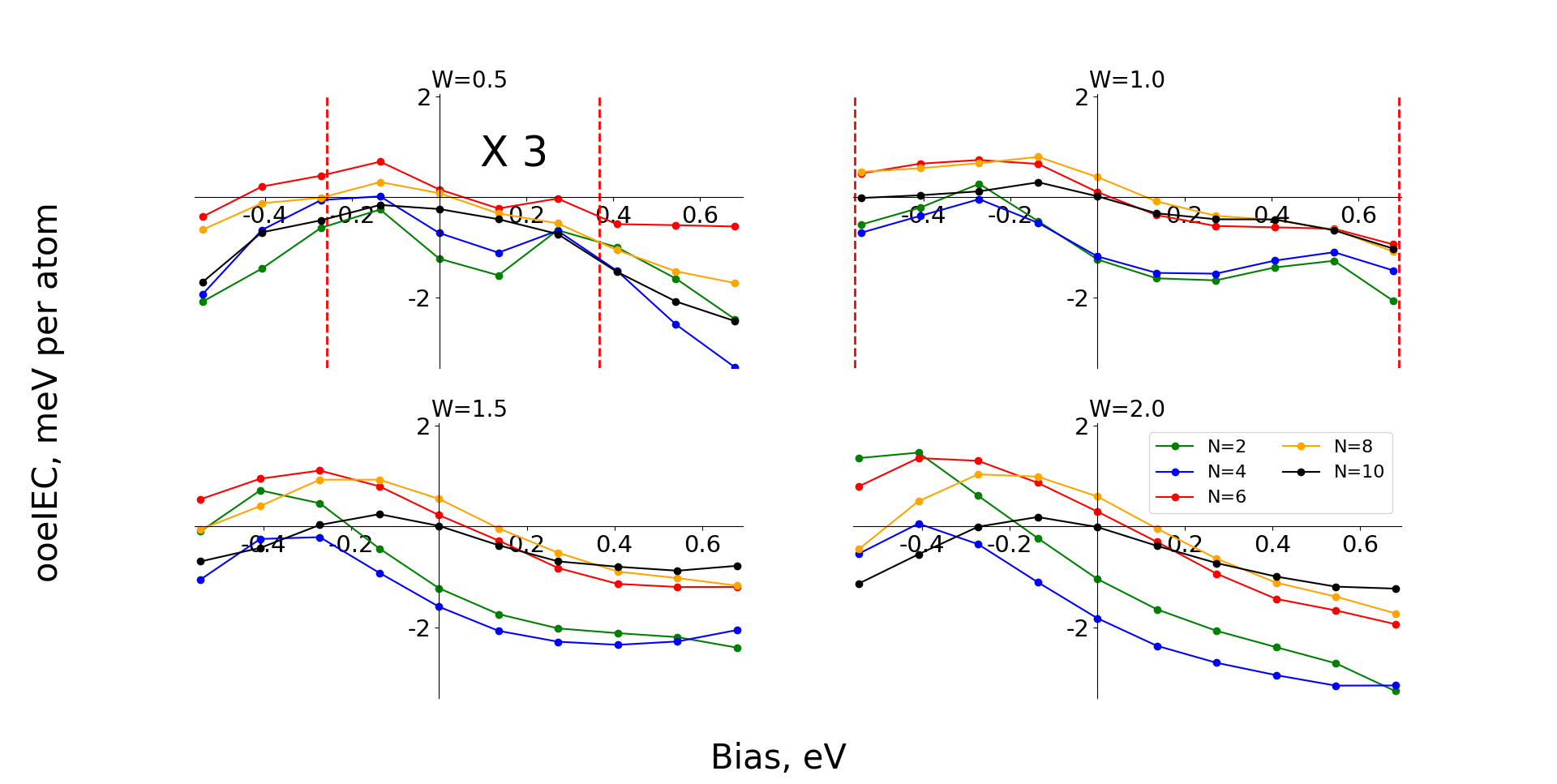}
    \caption{ooeIEC vs bias across the insulating gap for HG width parameter $W=0.5,1,1.5,2$ and barrier thickness $B=4$ throughout. For $W=0.5$ and $W=1$ the HG gap width is indicated by the vertical dashed lines. Note that for $W=0.5$ the ooeIEC data has been scaled by a factor of 3 for clarity.}
    \label{fig:fig 8}
\end{figure}

To study this, we modify our model $\text{Co}^{\downarrow}$ hopping $t_{sd}\rightarrow t_{sd} W$,
%\begin{equation}
%    \label{eqn:eqn 20}
%    \textbf{t}=
%    \left( \begin{array}{cc}
%      t_1   &  t_2\\
%      t_3   & t_4
%    \end{array}\right)
%    \rightarrow
%     \left( \begin{array}{cc}
%      t_1   &  Wt_2\\
%      Wt_3   & t_4
%    \end{array}\right)
%\end{equation}
where $W$ is a parameter which scales almost linearly with the minority Co HG width, 
 so that $W=1$ corresponds to no change, $W=0.5$ corresponds to roughly half the original width of $1.25$eV and so on. 
 FIG.~\ref{fig:fig 8} shows the ooeIEC versus applied bias for barrier thickness of $B=4$  and $W=0.5, 1.0, 1.5$ and $W=2.0$, corresponding to HG widths of $0.625$, $1.25$, $1.875$ and $2.5$eV respectively. 
 Note, that although the HG width is altered, the insulating section remains unchanged and so we continue to vary the bias across the range $-0.55\le eV \le 0.7$eV as before.

% \begin{figure}[h]
%     \centering
%     \includegraphics[width=0.7\linewidth]{test picture.png}
%     \caption{\textcolor{red}{alternative pic}}
% \end{figure}

\begin{figure}[h]
    \centering
    \includegraphics[width=0.7\linewidth]{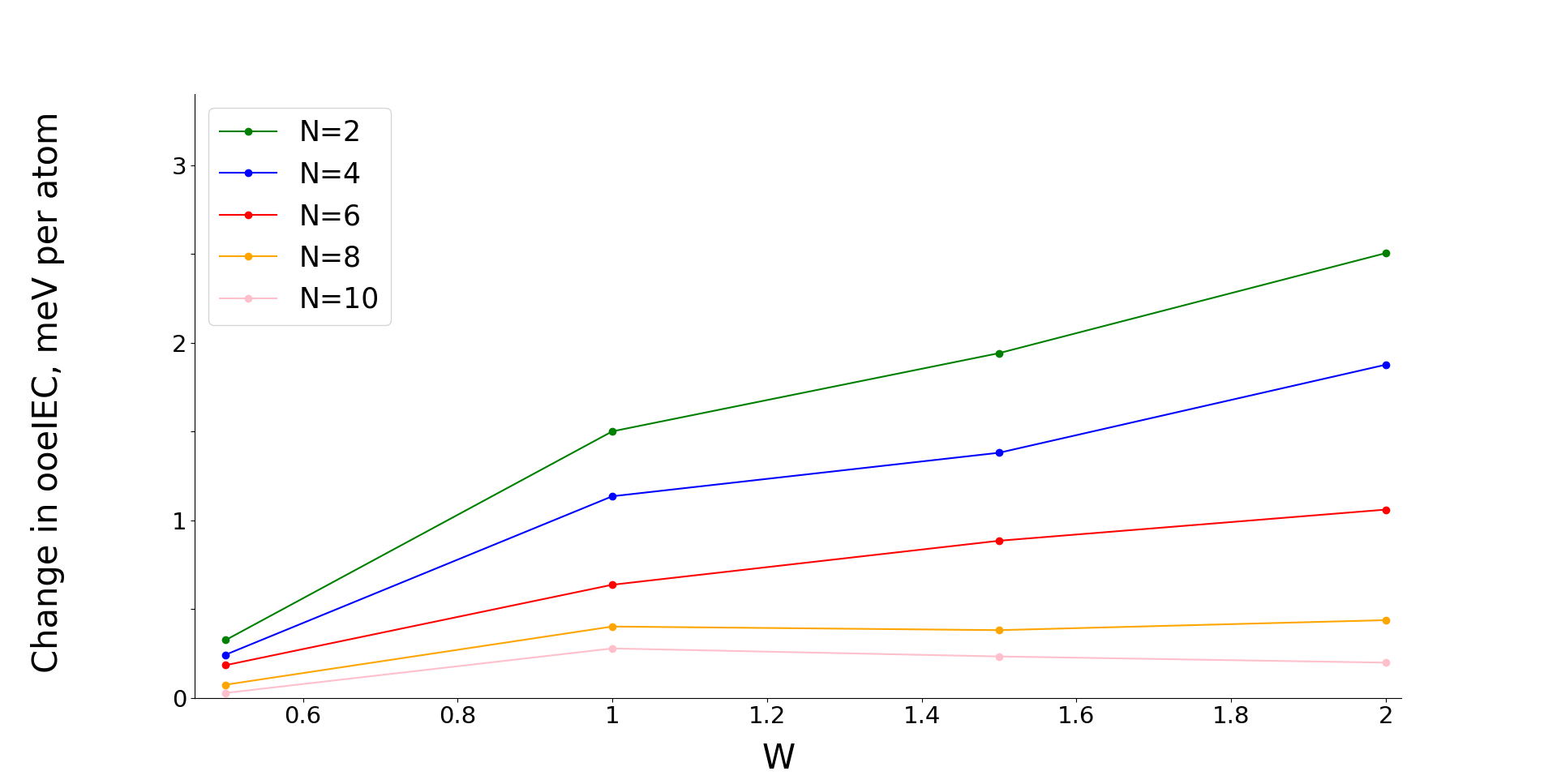}
    \caption{Maximum change in the ooeIEC for a single barrier system Cu/INS($4$)/FM(5)/Cu($N$)/FM(5)/Cu versus the FM HG width $W$.  $W=1$ corresponds to a Co FM.}
    \label{fig:fig 17}
\end{figure}

We observe that the predominant effect is that the amplitude of the ooeIEC decreases as the HG width decreases.  This is clearly seen in FIG.~\ref{fig:fig 17}, which shows the maximum change in the ooeIEC versus the FM HG width $W$ and we observe an approximate linear dependence on $W$.  

This is similar behaviour to the amplitude of the \emph{equilibrium} oscillatory IEC, which also decreases as the HG width decreases.  To see this we note that, by the stationary phase approximation \cite{mathon1997}, the amplitude $A$ of the equilibrium IEC, for fixed spacer thickness, is approximately given by
 \[    A \propto \frac{T}{N\sinh(\pi k_B T \psi')},\]
where $k_B$ is Boltzmann's constant, $T$ is temperature, and $\psi'={d\psi}/{d\epsilon}$, where $\psi$ is the phase of the first Fourier coefficient of the electron density of states.  In Ref.~\cite{UmerskiPhase}, we show that this phase change is fixed at approximately $2\pi$ as we move across the HG.  Hence $\psi'\sim 2\pi/(1.25W)$, where $1.25W$ eV is approximately the HG width, and so the amplitude $A \sim  W/N$, for the parameter ranges we consider.  FIG.~\ref{fig:fig 17} clearly illustrates such behaviour.

% The results of increasing the HG width and simultaneously the insulator gap width are shown in FIG. \ref{fig:fig 17}.  We observe that, for $W=1.5,\ 2$, the amplitude is somewhat enhanced but that the bias dependence weakens as $W$ increases further -- presumably because the insulator gap becomes so large as to cut the system from the right-hand lead.

% \textcolor{red}{I'm not sure what we conclude from this ... I can't tell if reducing/increasing the width enhances or reduces the effect.
% For discussion once all figures are in place.  \newline
% We conclude that the overall ooeIEC bias amplitude, when considered as a function of bias, is most strongly affected by the insulator thickness and band gap width but can also be affected by the HG width.}

\begin{figure}[h]
    \centering
    \includegraphics[width=0.7\linewidth]{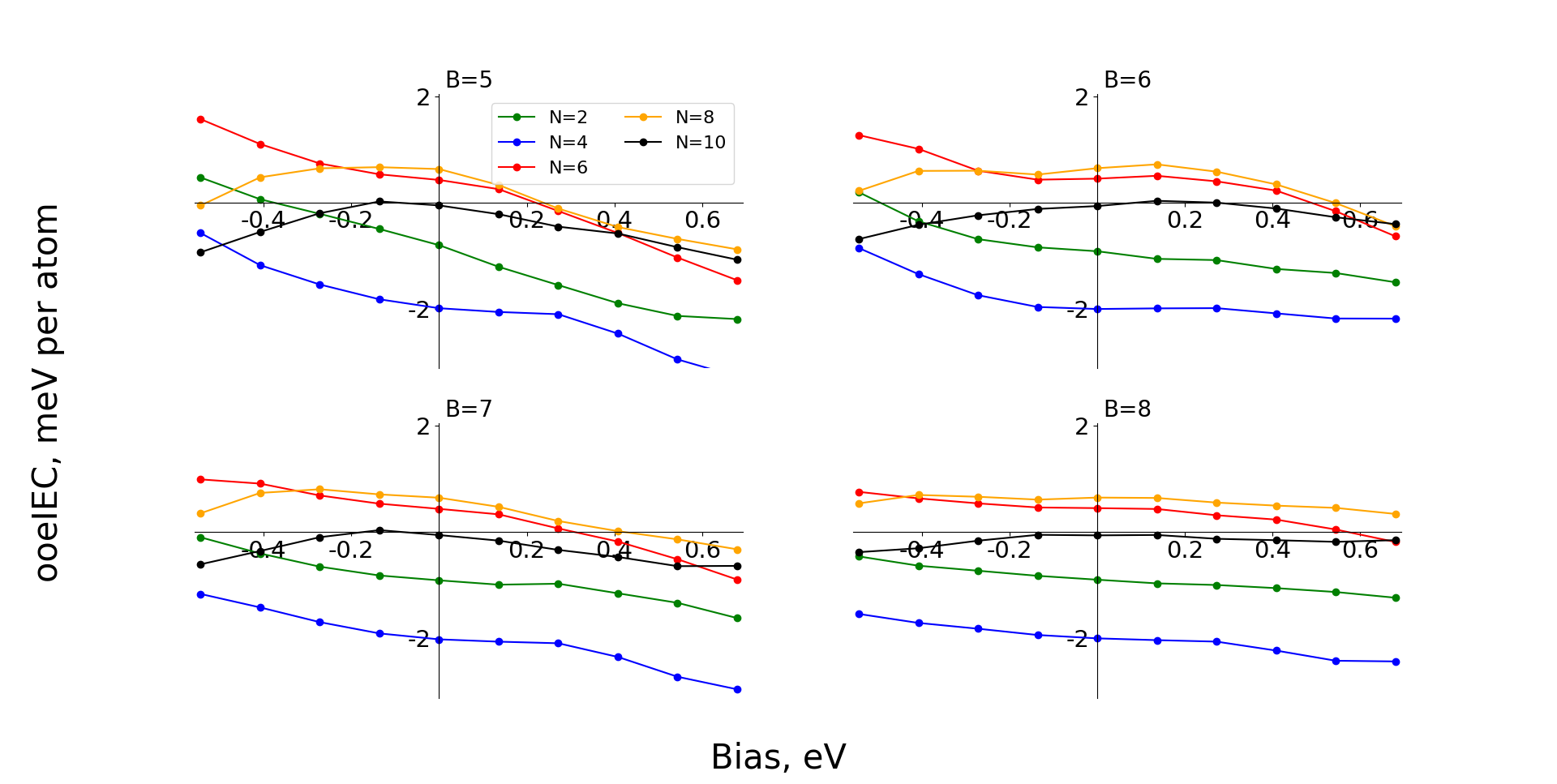}
    \caption{ooeIEC vs bias for FM HG width $W=3$, and barrier thicknesses $B=5, 6, 7, 8$.}
    \label{fig:fig 24}
\end{figure}

Conversely, the amplitude of the effect increases with increasing FM HG width leading to strong switching with thicker barriers and hence lower current densities.  Fig. \ref{fig:fig 24} shows the bias dependence of the ooeIEC with width parameter $W=3$, for barrier thicknesses $B=5, 6, 7$ and $8$ atomic layers.  A strong switching effect is observed up to $B=7$, which corresponds to a charge current density of no more than $10^7$ A/cm$^2$ -- an order of magnitude improvement over the $W=1$ case considered in the previous section.

\subsection{Double Barrier}
Section~\ref{sec:results}.A demonstrated that the bias dependence of the ooeIEC dies away quickly with increasing barrier thickness $B$ for a single barrier system. 
However, it is well known that electrons incident on a resonant tunnelling barrier, with momenta corresponding to the stationary states in the spacer well, will transmit with transmission coefficient unity.
We are therefore motivated to consider a double barrier system with insulating section of the form $\text{Co}^\downarrow$($B$)/Cu($TS$)/$\text{Co}^\downarrow$($B$). Here $TS$ is the thickness of the  spacer between the barriers, the total barrier thickness is $2B$, and we assume a constant linear potential drop across the barriers.  

% We would therefore expect that the total charge current for a DB system be comparable to that of a SB system, even with double the number of insulating layers. Charge current vs insulator thickness for a SB and DB system at two different bias's is shown in FIG. \ref{fig:fig 21} where we see that the charge current is indeed comparable.
%
%\textcolor{red}{Should linear drop for DB system be discussed in more detail? Maybe include another 1D potential schematic pic?}
%
%\begin{figure}[h]
%    \centering
%    \includegraphics[width=0.7\linewidth]{INS1 vs INS2 charge current.png}
%    \caption{Charge vs insulator thickness for SB and DB systems at $V=0.01,0.04$. Note that the total number of insulating layers for the DB system is twice that of the SB system. \textcolor{red}{B,b distinction unclear?}}
%    \label{fig:fig 21}
%\end{figure}
%
%\textcolor{red}{Furthermore, doubling the number of insulating layers is preferable for realistic devices given the reduced risk of dielectric breakdown.}
%
%\textcolor{red}{"symmetry" results may suggest $j_L$ and$ j_R$ more important than total charge current?}

\begin{figure}[h]
    \centering
    \includegraphics[width=0.7\linewidth]{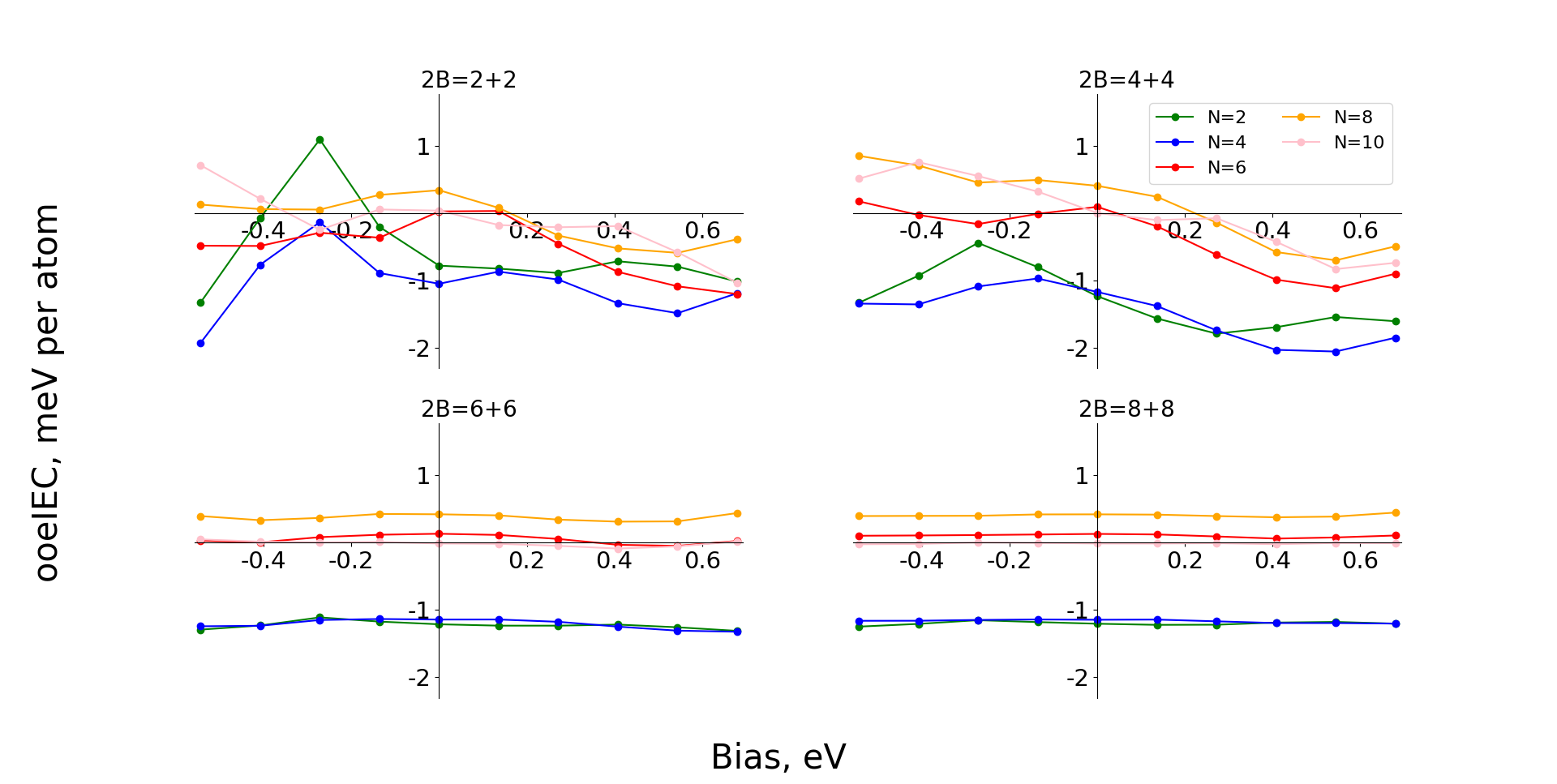}
    \caption{ooeIEC versus bias across the HG for a double 
    barrier system with barrier thicknesses $2B=2+2,4+4,6+6,8+8$. }
    \label{fig:fig 7}
\end{figure}

FIG. \ref{fig:fig 7} shows the ooeIEC versus bias for the model system \newline
 Cu/$\text{Co}^{\downarrow}$($B$)/Cu(6)/$\text{Co}^{\downarrow}$($B$)/Co(5)/Cu($N$)/Co(5)/Cu, with barrier thickness $2B=2+2,\ 4+4,\ 6+6, 8+8$.
% \begin{figure}[h]
%     \centering
%     \includegraphics[width=0.7\linewidth]{INS2 varying B delta j.png}
%     \caption{Maximum change in ooeIEC (as a function of bias) so DB system against $N$ for $2B=2+2,4+4,6+6,8+8$.}
%     \label{fig:fig 18}
% \end{figure}
This time we observe strong non-linear bias dependence for $2B=2+2$ and $2B=4+4$ with consistent change of sign of the ooeIEC for several spacer thicknesses $N$.   For thicker barriers $B \ge 6$, the bias dependence becomes too weak to overcome the equilibrium IEC.  For $B=4+4$ the charge current density is about $5\times10^{7}$A/cm$^2$ --- a factor of two improvement on the single barrier case.

\begin{figure}[h]
    \centering
    \includegraphics[width=0.7\linewidth]{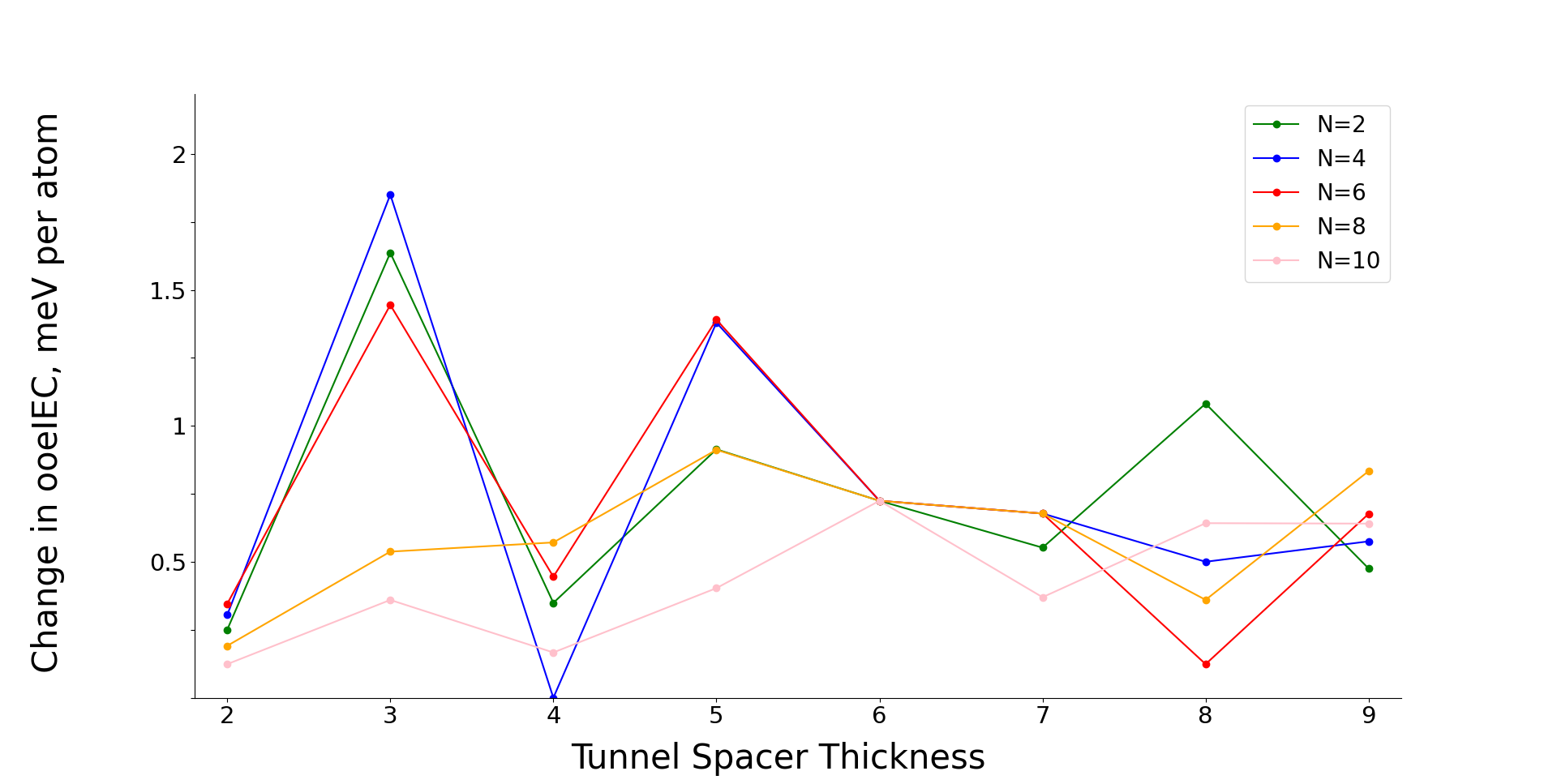}
    \caption{Maximum change in ooeIEC (with bias) for a double barrier system against the distance between the barriers.}
    \label{fig:fig 19}
\end{figure}

The spacer thickness between the barriers plays some role in determining the strength of the bias dependence on the ooeIEC. 
This is most clearly seen in FIG.~\ref{fig:fig 19} which shows the maximum change in ooeIEC (with bias) for the double barrier system against the distance between the barriers for a number of different spacer thicknesses $N$.  We observe strong oscillations in the magnitude of the effect, most likely due to constructive and destructive interference effects in the conductance between the barriers.

% \begin{figure}[h]
%     \centering
%     \includegraphics[width=0.7\linewidth]{INS2 for varying TS.png}
%     \caption{ooeIEC vs bias across the HG for an DB system and $TS=2,3,4,5$.}
%     \label{fig:fig 10}
% \end{figure}

%\textcolor{red}{From here on need discussing with Andrey and re-writing!!}
%
%Our $TS$ investigations also reveal an interesting symmetry between the IEC and STT terms. We consider the two IEC terms separately as functions of bias across the HG for $TS=2$ and 3 (with $N=4$) as shown in FIG. \ref{fig:fig 11}.
%
%\begin{figure}[h]
%    \centering
%    \includegraphics[width=0.7\linewidth]{INS2 sym vs antisym.png}
%    \caption{IEC and STT terms of ooeIEC vs bias across the HG for a DB system and $TS=2$ (left) and $TS=3$ (right). }
%    \label{fig:fig 11}
%\end{figure}
%
%The figures suggest that, for small charge current, the IEC and STT terms of the ooeIEC are out-of-phase leading to the lack of ooeIEC bias dependence. This phase coherency is destroyed as the charge current increases. In fact, this is straightforward to show algebraically from Equation (\ref{eqn:eqn 2}). In the limit as $j_L\to0$, we find that the two ooeIEC terms are related by a reflection in the horizontal axis and a rigid translation in the vertical direction causing them to be out-of-phase in this limit. \textcolor{red}{This demonstrates mathematically what we earlier deduced on physical grounds, namely the need for tunnelling across the barrier to give non-negligible charge current!!!}

\subsection{Amorphous Insulator}
In the previous subsections the insulating section was modelled assuming a perfect insulator growing epitaxially between the lead and exchange coupled junction.  However, the argument given in Section~\ref{sec:motiv} and depicted in FIG.~\ref{fig:fig 15}, of applying a finite bias across an insulator to access states throughout the HG of the FM, applies equally to epitaxially grown insulators or to amorphous insulators.  The insulator is merely a mechanism to maintain a finite bias, it is the exchange coupled junction which must be grown epitaxially in order to observe oscillatory IEC.

In this section we study the case where the insulating section is a model amorphous barrier,
consisting of a random mix of metallic and insulating atoms.  The metallic tight binding parameters are chosen to be the 2-band model potentials for Cu, given in Section~\ref{sec:potent}, while the insulating parameters are obtained by fitting a 2-band model to the empirical band structure of bulk MgO at the neck of Cu.  
 We set the Fermi energy to be at the centre of the band gap of MgO at the $\Gamma$-point, so that at the neck of Cu the band gap of MgO lies between $-3.9$ eV and 6.9 eV, ensuring that it is a good insulator for all energies within the HG of Co.
We find that the potentials presented in Table~\ref{tab:table2} replicate the band structure accurately.  

% \begin{table}
% \caption{Tight binding parameters for model MgO at the neck of Cu}\label{tab:table2}%
% \begin{tabular}{|c|c|c|c|c|c|}
% \hline
%   & $u_{ss}$ & $u_{sd}$ & $t_{ss}$ & $t_{sd}$ & $t_{dd}$\\
% \hline
% \ Cu\ & $-0.3263$ &\ 0.5001\ &\ 0.00558 \ &$-0.00242$\ & $\ 0.0065$ \\
% \hline
% \end{tabular}
% \end{table}
\begin{table}
\caption{Tight binding parameters for model MgO at the neck of Cu}\label{tab:table2}%
\begin{tabular}{|c|c|c|c|c|c|}
\hline
  & $u_{ss}$ & $u_{sd}$ & $t_{ss}$ & $t_{sd}$ & $t_{dd}$\\
\hline
\ Cu\ & $-4.4395$ & $6.8042$ & $0.0759$ & $-0.0329$ & $0.0884$ \\
\hline
\end{tabular}
\end{table}

Note that, it is necessary to employ a strong insulator like MgO in our calculations, as opposed to a weaker insulator such as the Co$^{\downarrow}$ parameters used in previous sections, because the latter requires too high a proportion of insulating:conducting atoms in order to create an effective barrier --- rendering the amorphous insulator almost perfectly crystalline.  Whereas using the strong insulating MgO potentials enables the creation of an amorphous insulator with a good mix of insulating:conducting atoms.

The amorphous insulator is modelled using a a 25-atom supercell arranged in a $5 \times 5$ grid in-plane, with a fixed proportion of Cu and MgO atoms placed at random locations within each layer in the growth direction.  All calculations of conductance and ooeIEC are averaged over 300 realisations to minimise the effects of random outliers and finite supercell.  See Ref.~\cite{durie22} for further details of the modelling procedure. 

\begin{figure}[h]
    \centering
    \includegraphics[width=0.7\linewidth]{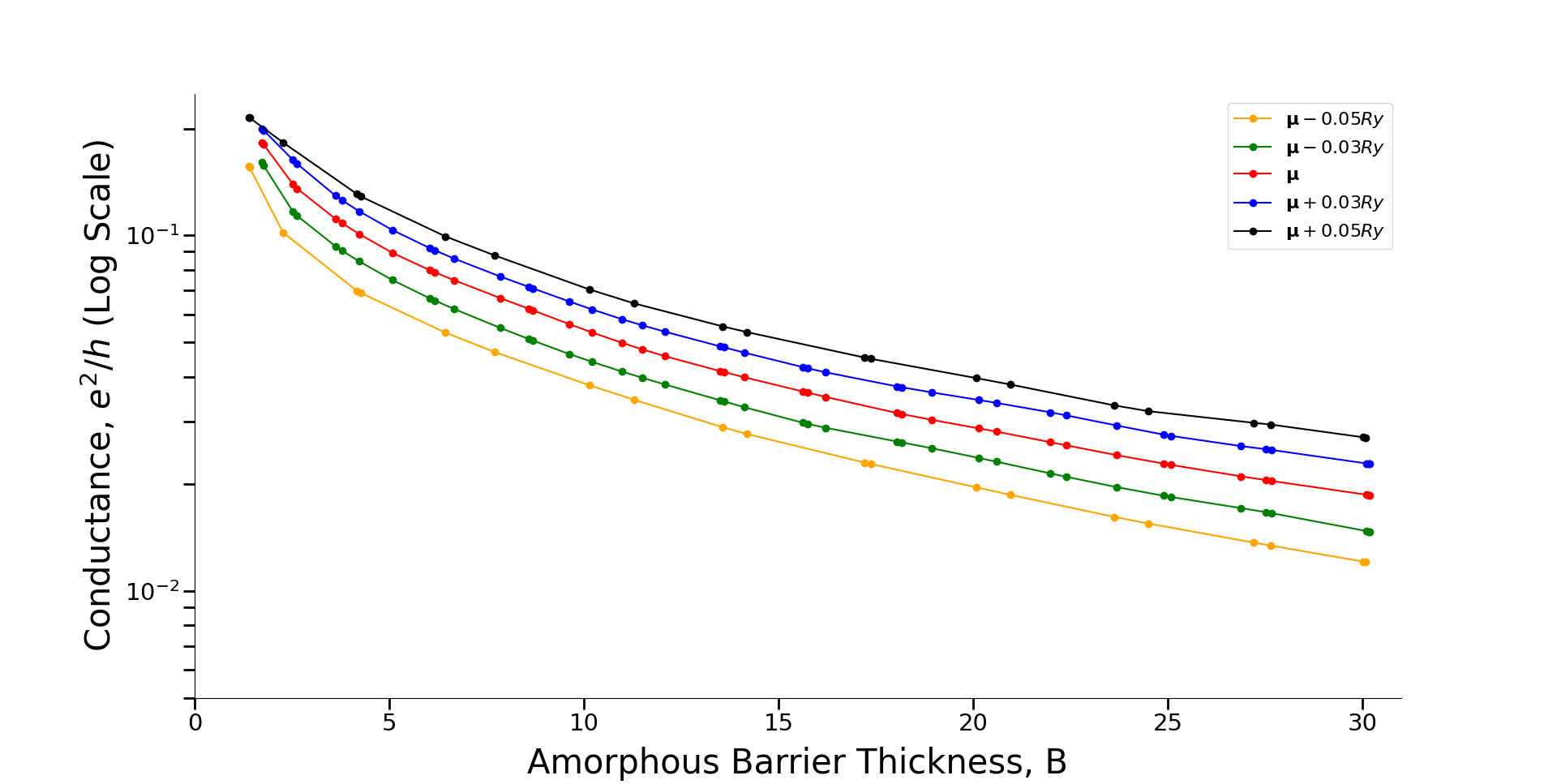}
    \caption{Conductance (on a logarithmic scale) of Cu/$\text{MgO}_{0.28}\text{Cu}_{0.72}(B)$/Cu as a function of the amorphous barrier thickness, for different energies.}
    \label{fig:fig 5}
\end{figure}

FIG.~\ref{fig:fig 5} shows the the conductance (on a logarithmic scale) as a function of the amorphous barrier thickness for a range of energies around the Fermi energy, for Cu/$\text{MgO}_{0.28}\text{Cu}_{0.72}$/Cu, where the amorphous insulator consists of 28\% MgO.
We observe that the conductance drops sharply for the first few planes, before settling into exponential decay after approximately 10 atomic planes -- indicating that the conductance mechanism is pure tunnelling and the amorphous layer is insulating for this range of energies. 

\begin{figure}[h]
    \centering
    \includegraphics[width=0.7\linewidth]{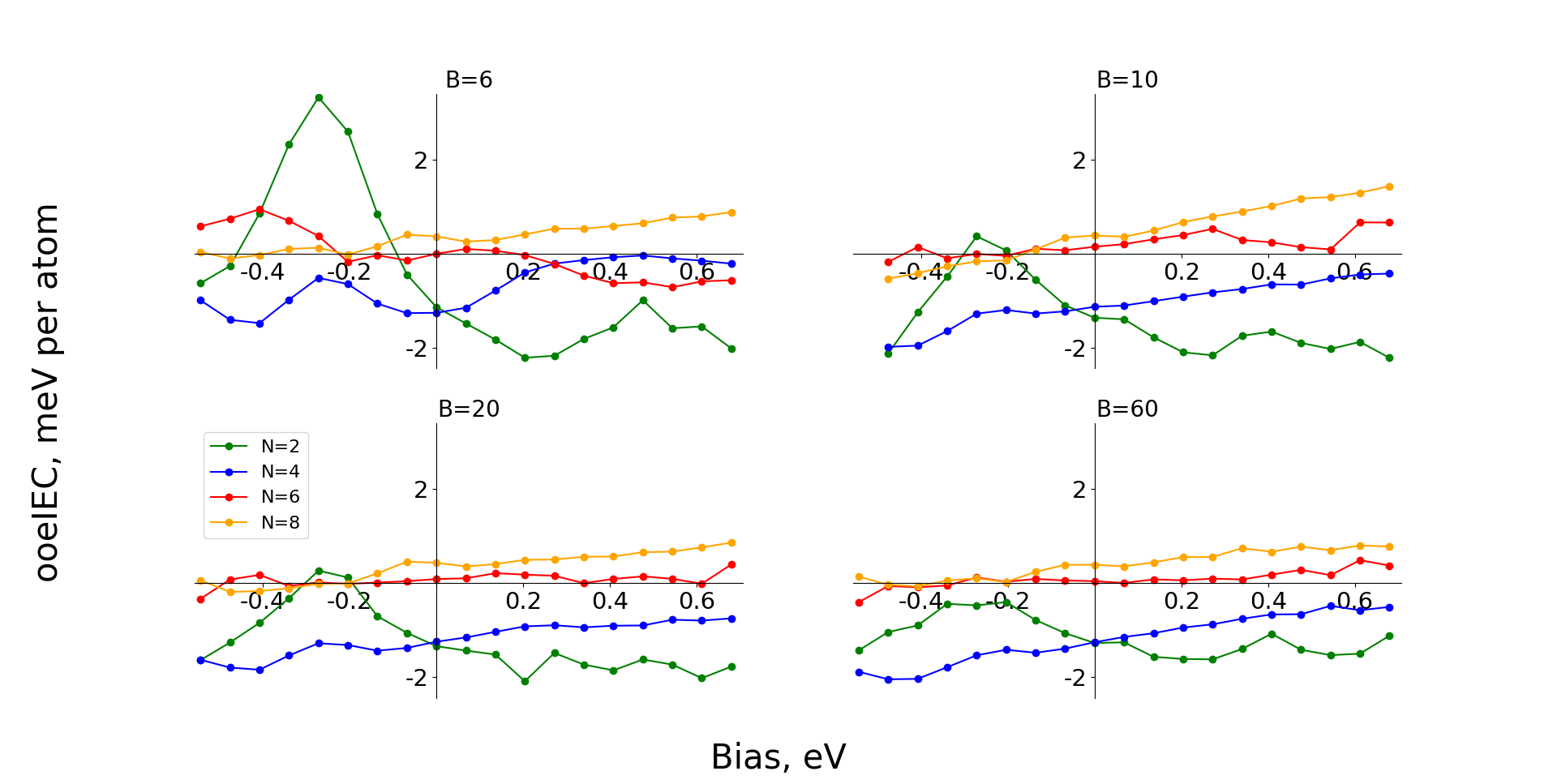}
    \caption{ooeIEC versus bias for Cu/$\text{MgO}_{0.28}\text{Cu}_{0.72}$($B$)/Co(10)/Cu($N$)/Co(10)/Cu with $N=4, 5, 6, 7$.  Top left $B=6$, top right $B=10$, bottom left $B=20$ and bottom right $B=60$. }
    \label{fig:fig 12}
\end{figure}

Using this amorphous barrier as the insulating section,
 FIG.~\ref{fig:fig 12} shows the bias dependence of the ooeIEC in the range of the FM HG, for a range of amorphous barrier thicknesses $B=6,10,20,60$.

We observe that, strong nonlinear behaviour is observed for thinner barriers $B$ and thinner spacer thickness $N$.  For thicker barriers and spacers the dependence is more linear.
Unlike crystalline insulators where bias dependence dies out after 8 or so layers, in this case some bias dependence remains even for the thickest amorphous insulators considered $B=60$.

%%%%%%%%%%%%%%%%%%%%%%%%%%%%%%%%%%%%%%%%%%%%%%%%%%%%%%%%%%%%%%%%%%%%%%%%%%%%%%%%%%%%%%%%

\section{Conclusion}

Quantum well oscillatory IEC occurs in an FM/NM/FM junction when there is confinement in the NM spacer for one of the electron spin states because it is trapped in a HG of the FM.
This is the case for most real systems exhibiting strong oscillatory IEC.
The presence of the HG ensures that the IEC is highly sensitive to the position of the Fermi-level within the gap, and in this communication we exploit this effect by applying a bias to engage states around the Fermi level in order to achieve magnetic switching. 

We apply an electrical bias to quantum well oscillatory IEC FM/NM/FM junctions, coupled to insulating sections across which the bias drops.  
 Using a 2-band model for Co as the FM, and Cu and the NM we have calculated the ooeIEC using a Landauer formalism for three different types of insulating section: a single barrier, a double barrier and an amorphous barrier.

In the single barrier case, we find that the ooeIEC dependence on bias dies away quickly with increasing barrier thickness, suggesting a strong correlation to the system conductance. 
For relatively weak barriers there is strong non-linear behaviour of the ooeIEC against bias, and the ground state of the system consistently switches between parallel and antiparallel magnetic alignment corresponding to switching. 
We also find that the strength of the ooeIEC effect is also dependent on the FM HG width $W$, and that its amplitude $A$ scales with that of the equilibrium IEC $A \sim W/N$, where $N$ is the distance between the FMs.
We conclude that the strongest effects for magnetic switching are achievable using FMs with larger HG widths and insulating sections corresponding to weaker barriers, and we estimate that materials with FM HGs widths of the order of $3.75$eV should achieve switching with current densities of less than $10^7$A/cm$^2$.

For double barrier insulating sections, the effect is similar to that of the single barrier case, except that the increased conductance due to resonant tunneling enhances the effect so that single barrier thickness $B$ is roughly equivalent to double barrier thickness $B+B$, with a reduction in switching current density of about a factor of two.

Finally, we have demonstrated that the effect does not depend on a high quality insulating section and that even for amorphous barriers, there can be strong dependence of ooeIEC on bias and magnetic switching can be achieved.

%%%%%%%%%%%%%%%%%%%%%%%%%%%%%%%%%%%%%%%%%%%%%%%%%%%%%%%%%%%%%%%%%%%%%%%%%%%%%%%%%%%%%%%%%%

\begin{acknowledgments}
We are grateful to the UK Engineering and Physical Sciences Research Council for financial support.
\end{acknowledgments}

\bibliography{PaperRefs}

\end{document}